\documentclass{ifacconf}

\usepackage{graphicx}      
\usepackage{natbib}        
\usepackage{amsmath}
\usepackage{tikz}
\usetikzlibrary{positioning,arrows.meta,shapes,calc}
\begin{document}
\begin{frontmatter}

\title{Benchmarking Sequential Feedback Optimization for Wind Farm Power Maximization\thanksref{footnoteinfo}} 

\thanks[footnoteinfo]{This work was partially supported by the NWO under research project
Online Optimization for Offshore Wind Farms.}

\author[First]{Shijie Huang} 
\author[First]{Sergio Grammatico} 

\address[First]{Delft Center of Systems and Control, Delft University of Technology, The Netherlands. (e-mail: \{s.huang-5, s.grammatico\}@ tudelft.nl).}

\begin{abstract}                
This paper benchmarks sequential feedback optimization (SFO) for wind farm power maximization using a medium-fidelity dynamic flow model. We compare SFO with two well-established approaches, adjoint-based economic model predictive control (AMPC) and extremum seeking control (ESC), under a common nine-turbine layout and identical operating constraints. The comparison focuses on steady-state power production and computational efficiency, both relevant for real-time implementation. The simulation results illustrate that SFO achieves higher steady-state power while preserving real-time feasibility, AMPC provides a better transient performance at a higher online computational cost and without guarantees of convergence to the steady-state optimum, and ESC offers a computationally inexpensive model-free baseline that may converge to locally optimal solutions. These results provide a practical reference for selecting wind farm control strategies and for designing scalable, real-time optimization methods.
\end{abstract}


\end{frontmatter}

\section{Introduction}
Wind power has become a key component of modern renewable energy systems. To reduce installation and maintenance costs, wind turbines are typically installed in close proximity to form wind farms. However, such arrangements lead to aerodynamic wake interactions, where upstream turbines reduce the available wind speed for downstream turbines, thereby lowering the total farm output \citep{boersma2017tutorial}. This has motivated intensive research on wind farm control strategies aimed at coordinating individual turbine settings to mitigate these interactions and enhance the collective performance.

Most existing wind farm control studies rely on simplified steady-state wake models such as the Jensen model \citep{jensen1983note}, Park model \citep{katic1987simple}, and the FLORIS framework \citep{gebraad2016wind}. These low-fidelity models describe wake interactions through analytical parametrizations and thus allow for developing efficient optimization methods, including sequential quadratic programming \citep{fleming2016wind}, sequential convex programming \citep{park2015cooperative}, and distributed proximal primal-dual algorithm \citep{annoni2019efficient}. While these approaches have demonstrated significant power improvement in low-fidelity simulations, their limited accuracy may lead to degraded optimization performance when evaluated with high-fidelity models \citep{kheirabadi2019quantitative}. This discrepancy highlights a research gap in effectively identifying the optimal steady-state power production strategies within the context of dynamic flow models with higher fidelity.

 Several studies have therefore focused on control-oriented dynamic models that more accurately capture wake interactions while maintaining computational efficiency \citep{boersma2018control, van2022adjoint}. Medium-fidelity flow solves such as the WindFarmSimulator (WFSim) \citep{boersma2018control} has been developed to resolve the unsteady wake evolution governed by the Navier-Stokes equations. Using such models, advanced optimization-based control methods, including adjoint-based economic model predictive control (AMPC) have been proposed to coordinate turbine  actions over finite prediction horizons \citep{vali2019adjoint, van2022adjoint}. These AMPC approaches can provide improved transient performance, although their real-time implementation remains computationally demanding and lack theoretical assurance for convergence to the optimal steady-state power production. In parallel, model-free control methods such as extremum seeking control (ESC) \citep{johnson2012assessment, kumar2023wind} and game-theoretic learning approaches \citep{marden2013model} have been applied to wind farms. These algorithms operate directly on power measurements, offering implementation simplicity and robustness to model uncertainties. Nevertheless, they often exhibit slow convergence and may get stuck in local optimal solutions \citep{ciri2017model}.  
 
 Recently, a sequential feedback optimization (SFO) method was proposed to steer a dynamic flow model to reach its steady-state operating point that maximizes total power production \citep{huang2025sequential}. The method replaces offline steady-state optimization with an online feedback-based process, iteratively refining control inputs based on measured power. Simulation studies have shown that SFO achieves significant power improvements compared to the baseline greedy controller. In this paper, we extend the previous study on SFO by further benchmarking it against AMPC and ESC methods using the same medium-fidelity dynamic model (WFSim). Our key contribution is a systematic comparison of the three methods, evaluated in terms of steady-state power output and computational efficiency. Under identical WFSim conditions, the simulation demonstrates that SFO provides a reasonable balance between steady-state power improvement and computational tractability. 
 
 The paper is organized as follows. Section~\ref{sec:model} presents the wind farm model. Section~\ref{sec:methods} describes the considered methods SFO, AMPC, and ESC. Section~\ref{sec:simulation} reports the comparative simulation results. Section~\ref{sec:conclusion} gives concluding remarks.


\section{Wind Farm Model}\label{sec:model}
When multiple wind turbines operate in the same area, the wind flowing through upstream wind turbines will generate wakes with reduced wind speed and increased turbulence. These wakes interact with downstream turbines, leading to significant power losses and consequently reducing the overall wind farm efficiency. The objective of wind farm control strategies is therefore to enhance the power production by mitigating the wake losses. Achieving this requires an aerodynamic model that can capture the dominant flow physics. In this section, we briefly introduce the turbine power function and the employed wake interaction model.
\subsection{The Turbine Model}
Consider a wind farm composed of $N$ wind turbines, each with rotor radius $R$. For turbine $i$, let $a_i$ denote the axial induction factor and $\gamma_i$ the yaw angle. While yaw angle is a control variable that can be directly manipulated, the induction factor is indirectly adjusted by controlling the blade pitch angle and generator torque of the wind turbine \citep{boersma2017tutorial}.

Under actuator disk theory, the power and thrust generated by turbine $i$ are expressed as \cite[Eq. (5)]{boersma2017tutorial}
\begin{align}
P_i(a_i, \gamma_i, v_i) &= \frac{1}{2}\rho\pi R^2 C_P(a_i, \gamma_i)v_i^3,\\
F_i(a_i, \gamma_i, v_i) &= \frac{1}{2}\rho\pi R^2 C_T(a_i, \gamma_i)v_i^2,
\end{align}
where $v_i$ is the local velocity experienced by the turbine and $C_P$, $C_T$ represent the power and thrust coefficients defined by \cite[Eq. (6)]{boersma2017tutorial}
\begin{align}
C_P(a_i, \gamma_i) &= 4a_i(1-a_i)^2\cos^3(\gamma_i),\\
C_T(a_i, \gamma_i) &= 4a_i(1-a_i)\cos^2(\gamma_i).
\end{align}

Since the local velocity $v_i$ is influenced by the control variables of other turbines, coordinated control is necessary to optimize the collective objective. Previous studies have shown that yaw misalignment is an effective strategy to improve total power by deflecting the wake away from downstream turbines \citep{doekemeijer2020closed}. 
\subsection{The Wake Interaction Model}
As in \cite{vali2019adjoint} and \cite{huang2025sequential}, we employ WFSim, a dynamic flow model, to capture the flow dynamics among turbines. WFSim numerically solves the two-dimensional incompressible Navier-Stokes equations at hub height, computing both the longitudinal and lateral velocity components. After a spatial discretization on a staggered grid, the discrete-time nonlinear state-space representation can be written as: 
\begin{equation}\label{WFSim_model}
\begin{cases}
E(X_k)X_{k+1} = AX_k + b(X_k, \nu_k, \gamma_k)\\
y_{i,k} = P_i(\nu_{i,k}, \gamma_{i,k}, X_k), \quad i = 1, \ldots, N,
\end{cases}
\end{equation}
where $X_k =(u_k, v_k, p_k)$ represents the state vector, with $u_k$ and $v_k$ denoting the discretized flow velocities in the streamwise and lateral directions, and $p_k$ the pressure field. The nonlinear term $b(X_k, \nu_k, \gamma_k)$ includes the actuator disk forces and the coefficient matrix is given by
\begin{equation}
E(X_k) = \begin{bmatrix}
A_u(u_k, v_k) & 0 & B_1 \\
0 & A_v(u_k, v_k) & B_2 \\
B_1^T & 2B_2^T & 0
\end{bmatrix},
\end{equation}
\textcolor{black}{where $A_u,A_v,B_1,B_2$ are the WFSim discretization matrices associated with the velocity and pressure variables.} The control variables are the disk-based thrust coefficients and yaw angles, denoted as
\begin{equation}
\nu_k = [C'_{T,i,k}]_{i=1}^N, \qquad \gamma_k = [\gamma_{i,k}]_{i=1}^N.
\end{equation}
\textcolor{black}{The prime in $C'_T$ distinguishes the WFSim disk-based thrust coefficient from the conventional thrust coefficient $C_T$ used in actuator-disk theory. Under the ideal actuator-disk relation without yaw misalignment, $C'_T=4a/(1-a)$. The WFSim turbine model also uses calibration parameters $c_p$ and $c_f$ when mapping $C'_T$ to power production and actuator-disk forcing. For a more detailed description of the numerical implementation and validation of WFSim, the reader is referred to \cite{boersma2018control}.}

\section{Methods Overview}\label{sec:methods}

Various wind farm control approaches have been proposed to maximize the total power production under wake interaction described by a dynamic model. This section briefly reviews three representative strategies considered in this study: sequential feedback optimization (SFO), adjoint-based economic model predictive control (AMPC), and extremum seeking control (ESC).
\subsection{Sequential Feedback Optimization}

The SFO algorithm has been proposed in \cite{huang2025sequential} to iteratively drive the dynamic flow model toward its steady state that maximizes total  power production. The corresponding steady-state optimization problem can be written as
 \begin{equation}\label{power_max}
\begin{split}
\min\limits_{\nu,\gamma}\  &J(\nu,\gamma,y) := \left(\frac{\mathbf{1}_N^T y - P^{\text{ref}}}{P^{\text{ref}}}\right)^2 + \mu\|\nu\|^2 + \mu_\gamma\|\gamma\|^2 \\
\text{s.t.} \ \  & y = h(\nu, \gamma),\\
& \nu_i \in [\nu_{\min}, \nu_{\max}],\ \forall i = 1, \ldots, N, \\
& \gamma_i \in [\gamma_{\min},\gamma_{\max}],\ \forall i = 1, \ldots, N.
\end{split}
\end{equation}
Here $h(\nu,\gamma)$ denotes the steady-state input-output map of WFSim, $P^{\text{ref}}$ is a given reference total power, and $\mu$ as well as $\mu_\gamma$ are small regularization coefficients that prevent excessive control actions and improve numerical stability.

Since the exact mapping $h(\nu,\gamma)$ of WFSim is implicit and high-dimensional, the explicit sensitivity $\nabla h(\nu,\gamma)$ is unavailable. SFO uses the real-time power measurements to replace the steady power and estimates the sensitivity through sequential local linearization of WFSim. Specifically, at iteration $k$, the approximate sensitivities are computed as 
\begin{align}
\widetilde{\nabla_\nu h}(X_k,\nu_k,\gamma_k) &= C_k\,(E(X_k)-\mathcal{A}_k)^{-1}B_{1,k} + D_{1,k},\\
\widetilde{\nabla_\gamma h}(X_k,\nu_k,\gamma_k) &= C_k\,(E(X_k)-\mathcal{A}_k)^{-1}B_{2,k} + D_{2,k},
\end{align}
where the system matrices $\mathcal{A}_k,B_{1,k},B_{2,k},C_k,D_{1,k},D_{2,k}$ are obtained from the Jacobians of the WFSim model and actuator-disk relations (see \cite{huang2025sequential} for details). These sensitivity matrices are updated at each iteration using the latest flow states and control inputs, enabling the SFO algorithm to adaptively track the evolving wind farm dynamics. The control update laws are then given by  
\begin{align}
\nu_{k+1} &= \mathrm{proj}_{[\nu_{\min}, \nu_{\max}]}\Big(\nu_{k} - \alpha_\nu\Big(\nabla_\nu J\left(\nu_{k}, \gamma_k, P_{k}\right)\notag\\
&\quad + \widetilde{\nabla_\nu h}\left(X_k, \nu_{k}, \gamma_{k}\right)^{\top}\nabla_yJ\left(\nu_{k}, \gamma_k, P_{k}\right)\Big)\Big)\\ \label{SFO-v}
\gamma_{k+1} &= \mathrm{proj}_{[\gamma_{\min}, \gamma_{\max}]}\Big(\gamma_{k} - \alpha_{\gamma}\Big(\nabla_{\gamma} J\left(\nu_{k}, \gamma_k, P_{k}\right)\notag\\
&\quad + \widetilde{\nabla_\gamma h}\left(X_k, \nu_{k}, \gamma_{k}\right)^{\top}\nabla_yJ\left(\nu_{k}, \gamma_k, P_{k}\right)\Big)\Big)\\ \label{SFO-gamma}
\end{align}
where $\alpha_\nu, \alpha_\gamma > 0$ are the step size and $P_k = \mathbf{1}_N^{\top} y_k$ is the total measured power. 

The SFO method inherits the robustness of feedback optimization scheme \citep{colombino2019online, hauswirth2024optimization}, compensating modelling errors through direct power measurements. In addition, through successive re-linearization, SFO adapts toward an efficient steady state while maintaining feasible inputs. Under smoothness and Lipschitz assumptions on $h$, convergence to a neighborhood of the optimal steady state is established in \cite[Theorem 3.6]{huang2025sequential}.

\subsection{Adjoint-Based Model Predictive Control}

When a sufficiently accurate flow model is available, the AMPC framework is a widely used method for improving wind farm power maximization \citep{vali2019adjoint, van2022adjoint}. At each time step, an optimal control problem is solved to maximize the predicted cumulative power output over a finite horizon $N_p$:
\begin{equation}\label{EMPC-objective}
\begin{split}
	\max\limits_{\tilde{X},\tilde{\nu},\tilde{\gamma}}\ &\mathcal{J}(\tilde{X},\tilde{\nu},\tilde{\gamma}):= \sum_{k=1}^{N_p}\sum_{i=1}^N P_{i,k}(X_k,\nu_k, \gamma_k)\\
	\text{s.t.}\ \  & E(X_{k-1})X_k - AX_{k-1} - b(X_{k-1},\nu_{k-1}, \gamma_{k-1}) = 0,\\
	& \nu_k\in [\nu_{\min},\nu_{\max}],\ k=1,\dots,N_p,\\
	& \gamma_k\in [\gamma_{\min}, \gamma_{\max}], \ k=1,\dots,N_p.
\end{split}
\end{equation}
Here $\tilde{X} = (X_1, \dots, X_{N_p})^{\top}$ denotes the predicted flow states, while $\tilde{\nu} = (\nu_1,\dots,\nu_{N_p})^{\top}$ and $\tilde{\gamma} = (\gamma_1,\dots,\gamma_{N_p})^{\top}$ represent the sequences of control inputs. The large-scale optimization problem is typically solved using adjoint methods to efficiently compute the gradient of the cumulative power with respect to all control inputs (see \cite{vali2019adjoint} for the details). A gradient ascent step is then performed as
\begin{align}
\tilde{\nu}^{(t+1)} &= \tilde{\nu}^{(t)} + \alpha_\nu \frac{\partial\mathcal{J}}{\partial \tilde{\nu}}\\
\tilde{\gamma}^{(t+1)} &=  \tilde{\gamma}^{(t)} + \alpha_{\gamma} \frac{\partial\mathcal{J}}{\partial \tilde{\gamma}},
\end{align}
where $\mathcal{J}$ is the economic objective in \eqref{EMPC-objective}. \textcolor{black}{In the numerical implementation, a control horizon $N_c\leq N_p$ can be used for the optimized input sequence. After each optimization update, the first $N_r$ control inputs are applied to the plant model. The previous input sequence is then shifted forward by $N_r$ steps and the last input is repeated to initialize the next optimization update.} This implementation scheme is known as real-time iteration (RTI) in MPC literature \citep{diehl2005real, gros2020linear} (hereafter we write AMPC (RTI)).

Unlike SFO, which explicitly seeks the steady state that maximizes total power, AMPC (RTI) optimizes cumulative power over a finite horizon. The resulting closed-loop solution may differ from the optimal steady state. However, AMPC (RTI) anticipates future wake interactions through the prediction horizon and often provides a fast initial transient response. Closed-loop stability of economic MPC has been established under dissipativity and regularity assumptions \citep{amrit2011economic, angeli2011average}. 

\subsection{Extremum Seeking Control}

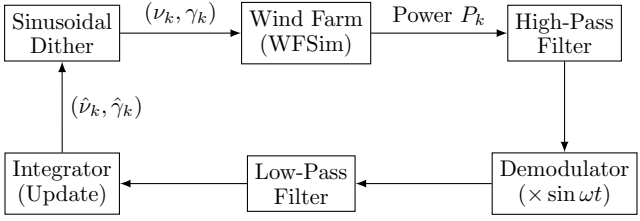
\begin{figure}[!t]
\centering
\begin{tikzpicture}[auto, node distance=1.5cm,>=latex,scale=0.85, every node/.style={scale=0.85}]
\node [draw, rectangle, align=center] (sum) {Sinusoidal \\ Dither};
\node [draw, rectangle, right=1.6cm of sum, align=center] (plant) {Wind Farm \\ (WFSim)};
\node [draw, rectangle, right=1.8cm of plant, align=center] (hpf) {High-Pass \\ Filter};
\node [draw, rectangle, below=1.2cm of hpf, align=center] (dem) {Demodulator \\ ($\times \sin \omega t$)};
\node [draw, rectangle, left=1.8cm of dem, align=center] (lpf) {Low-Pass \\ Filter};
\node [draw, rectangle, left=1.68cm of lpf, align=center] (int) {Integrator \\ (Update)};
\draw [->] (sum) -- node[above]{$(\nu_k,\gamma_k)$} (plant);
\draw [->] (plant) -- node[above]{Power $P_k$} (hpf);
\draw [->] (hpf) -- (dem);
\draw [->] (dem) -- (lpf);
\draw [->] (lpf) -- (int);
\draw [->] (int) -- node[right]{$(\hat{\nu}_k,\hat{\gamma}_k)$} (sum);
\end{tikzpicture}
\caption{Schematic structure of the ESC loop.}
\label{fig:esc_diagram}
\end{figure}

Extremum seeking control (ESC) represents a model-free alternative that optimizes performance directly from measured power without requiring an explicit model of the flow field \citep{johnson2012assessment, kumar2023wind}. The key idea is to inject a small sinusoidal perturbation into the control input and feed it into the plant model. The resulting measured power is processed by demodulation and low-pass filtering to estimate the gradient of the objective function, denoted by $\widehat{\nabla_\nu J}$ and $\widehat{\nabla_\gamma J}$. The control variables are then updated in the ascent direction,
\begin{align}
\hat{\nu}_{k+1} &= \hat{\nu}_k + \alpha_\nu\,\widehat{\nabla_\nu J},\\
\hat{\gamma}_{k+1} &= \hat{\gamma}_k + \alpha_\gamma\,\widehat{\nabla_\gamma J}.
\end{align}
The algorithmic structure of ESC is illustrated in Fig. \ref{fig:esc_diagram}. 

The ESC method is robust to model uncertainty, and requires extremely low computational cost, but convergence is typically slower and limited to local optima. Specifically, if the perturbation amplitudes and frequencies are properly chosen, ESC is guaranteed to converge to a neighborhood of a local optimum \citep{nevsic2009extremum}.

\section{Simulation Results}\label{sec:simulation}

\begin{figure}
\begin{center}
\includegraphics[width=8cm]{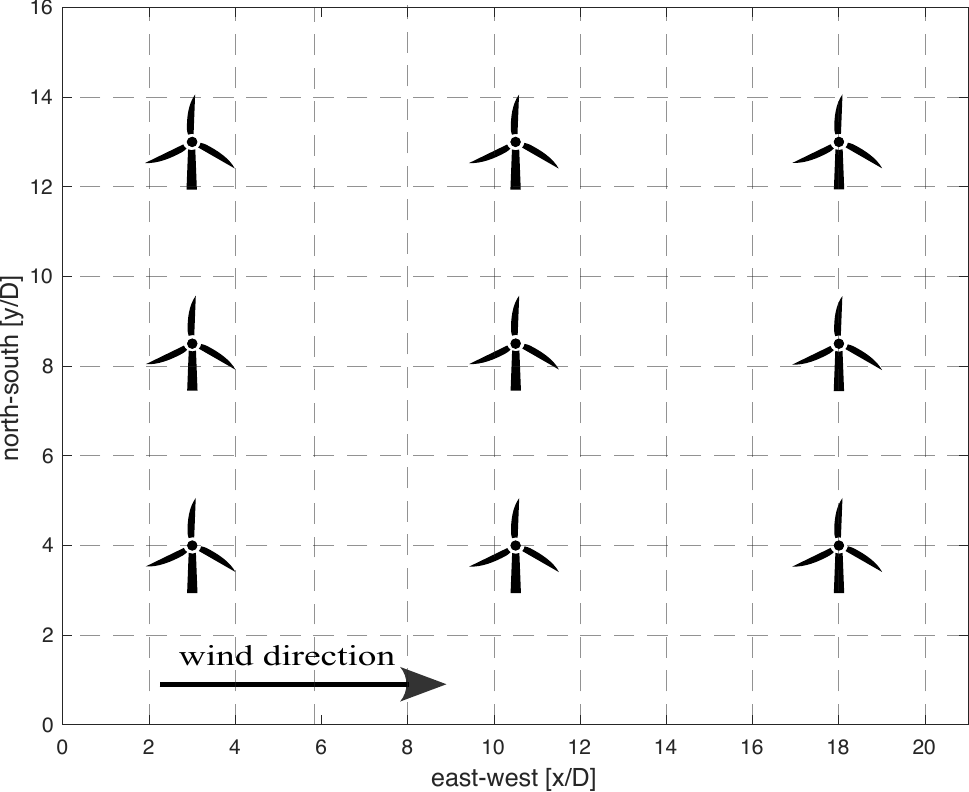}    
\caption{\textcolor{black}{Schematic wind farm layout with turbine positions normalized by $D$}.} 
\label{fig:layout}
\end{center}
\end{figure}

This section evaluates the three wind farm power maximization methods introduced in the previous section using a wind farm consisting of $9$ NREL 5MW reference turbines in a $3\times 3$ square grid, as shown in Fig. \ref{fig:layout}. The inter-turbine spacing follows the configuration adopted in \cite{boersma2018control}, ensuring significant wake interaction. Specifically, \textcolor{black}{The rotor diameter is denoted by $D = 126.4\ \mathrm{m}$, and the streamwise and spanwise turbine spacings are $5 D$ and $3 D$, respectively.}  All methods are simulated using the same flow model (WFSim) with a $2518.8\ \mathrm{m}\times 1558.4\ \mathrm{m}$ domain discretized on a $50\times 25$ staggered grid and sampling time $\Delta t=1\,\mathrm{s}$. The inflow conditions are assumed steady and uniform with $u_b = 8 \mathrm{m/s}$ and $v_b = 0 \mathrm{m/s}$. In the simulations below, the WFSim turbine-force and power parameters are set to $c_f=1.9$ and $c_p=0.99$, respectively, following the employed WFSim configuration.

\subsection{Total Power Production}

\textcolor{black}{All controllers are evaluated under the same WFSim model, greedy-flow initialization, and actuator bounds. The disk-based thrust coefficient for each turbine is constrained to $C'_T \in [0.5, 2.5]$,  and the yaw angle is limited to $\gamma \in [-30^{\circ}, 30^{\circ}]$. Under the ideal actuator-disk relation $C'_T=4a/(1-a)$, the Betz-optimal induction $a=1/3$ corresponds to $C'_T=2.0$. Therefore, the selected interval therefore allows mildly high-induction operating points, with the upper bound corresponding to $a=0.385$.
Similar high-induction settings have also been considered in wind farm control studies \citep{cossu2021wake, munters2016effect}.}

\textcolor{black}{Before applying each controller, WFSim is first run for $6000\,\mathrm{s}$ under greedy control (i.e., $C_{T}'^{\mathrm{greedy}} = 2$ and $\gamma^{\mathrm{greedy}} = 0^{\circ}$ for all turbines) to obtain a steady flow field. The resulting flow field and greedy control inputs are used as the common initial condition for all methods. Thus, $t=0$ in the following figures denotes the start of the optimization after the same flow field is developed. This common initialization avoids introducing an additional model-dependent initial guess.}

\textcolor{black}{The controller parameters were selected through preliminary simulations under the common benchmark setup to obtain stable closed-loop trajectories under the imposed bounds. For the SFO formulation in \eqref{power_max}, we set the total power reference to $P^{\mathrm{ref}} = 18\,\mathrm{MW}$ and choose regularization parameters $\mu = 2.8\times 10^{-3}$ and $\mu_{\gamma} = 1.5\times 10^{-5}$. For AMPC (RTI), we use a prediction horizon and control horizon of $N_p=N_c=200$ time steps and a receding step of $N_r=1$. At each sampling step, one projected adjoint-gradient RTI update is performed. For ESC, the dither amplitudes are $A_\gamma=2^\circ$ and $A_{C'_T}=0.05$, the dither frequencies are $\omega_\gamma\in[0.006,0.014]\,\mathrm{rad/s}$ and $\omega_{C'_T}\in[0.012,0.028]\,\mathrm{rad/s}$, the high-pass and low-pass filter cut-off frequencies are $\omega_{\mathrm{hpf}}=\omega_{\mathrm{lpf}}=0.002\,\mathrm{s}^{-1}$, and the adaptation gains are $k_\gamma=5\times 10^{-3}$ and $k_{C'_T}=5\times10^{-4}$. }

\begin{figure}
\begin{center}
\includegraphics[width=8.8cm]{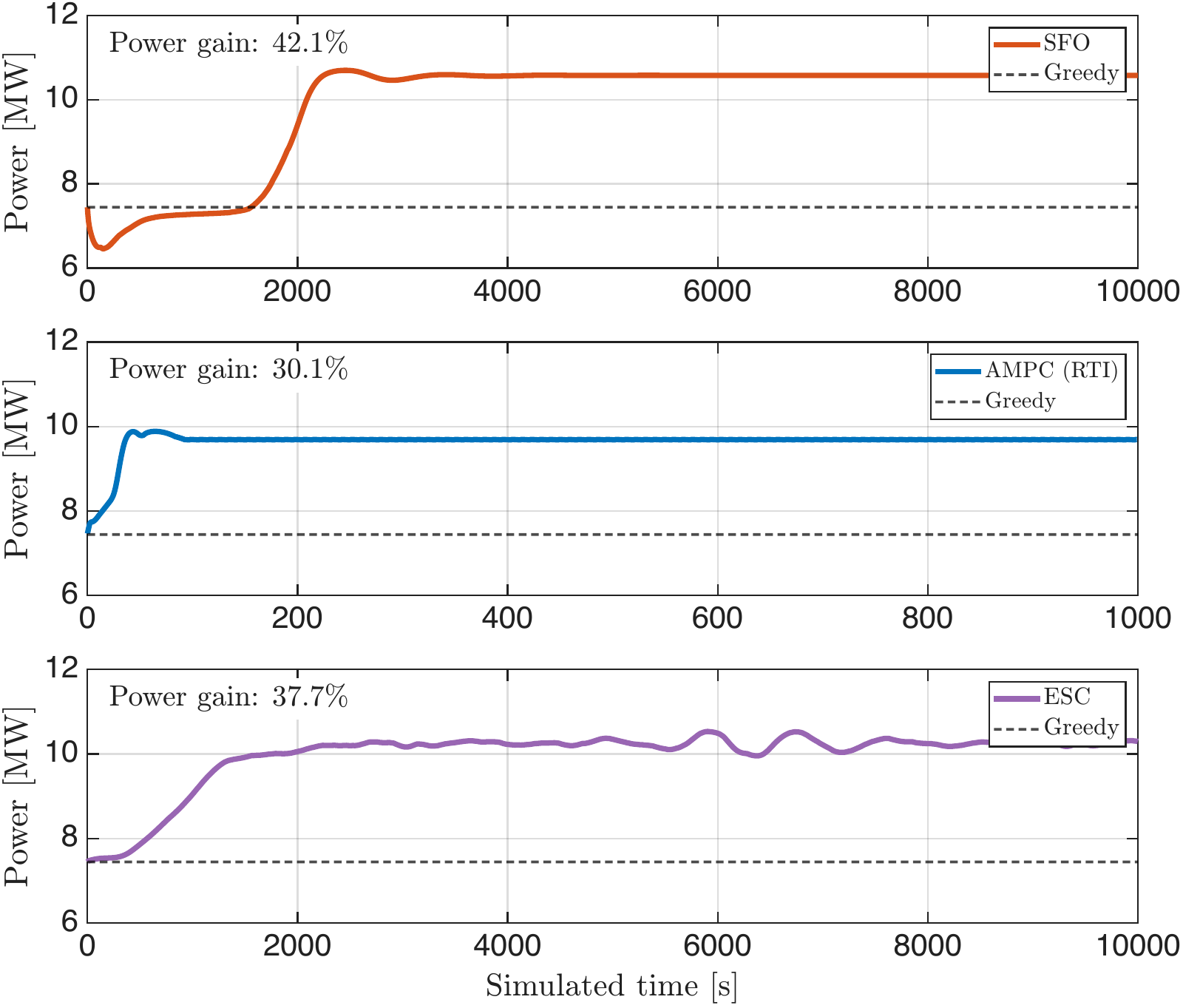}    
\caption{Total farm power vs. simulated time for SFO, AMPC (RTI), and ESC. \textcolor{black}{The AMPC trajectory is reported over a $1000\,\mathrm{s}$ window, which exceeds the characteristic wake-advection time scale of the layout.}} 
\label{fig:power_separate}
\end{center}
\end{figure}

Fig. \ref{fig:power_separate} illustrates the trajectories of the total farm power for SFO, AMPC (RTI), ESC, together with the greedy baseline. \textcolor{black}{The percentages reported in the panels indicate the final power gain relative to the greedy baseline.} In the considered steady-inflow case, SFO achieves the highest final power gain. AMPC (RTI) shows a fast initial transient response but reaches a lower power level. ESC achieves a comparable improvement but exhibits persistent oscillations caused by its continuous sinusoidal excitation signal, which is standard for extremum seeking schemes that estimate local gradients in real time without a model \citep{kumar2023wind}.

The observed differences might be explained by the underlying control objectives. \textcolor{black}{SFO explicitly targets a regularized steady-state operating point, which causes it to gradually adjust control actions toward a better steady configuration while avoiding unnecessarily aggressive inputs}. AMPC (RTI), by contrast, utilizes its predictive capability to realize rapid early gains but settles at a lower power level in this case. This may be related to the finite-horizon formulation that optimizes cumulative rather than asymptotic power. \textcolor{black}{Longer horizons, terminal costs, or model-based initial guesses such as those obtained from FLORIS may further improve AMPC performance, but a systematic study of these MPC-specific design choices is outside the scope of this benchmark.} ESC, although robust to modelling errors, tends to settle near a  local optimum because it relies purely on measurement feedback without exploiting model information. Overall, SFO combines steady-state efficiency with stable convergence, while AMPC (RTI) and ESC emphasize transient response and computational simplicity, respectively.

\begin{figure}
\begin{center}
\includegraphics[width=8.8cm]{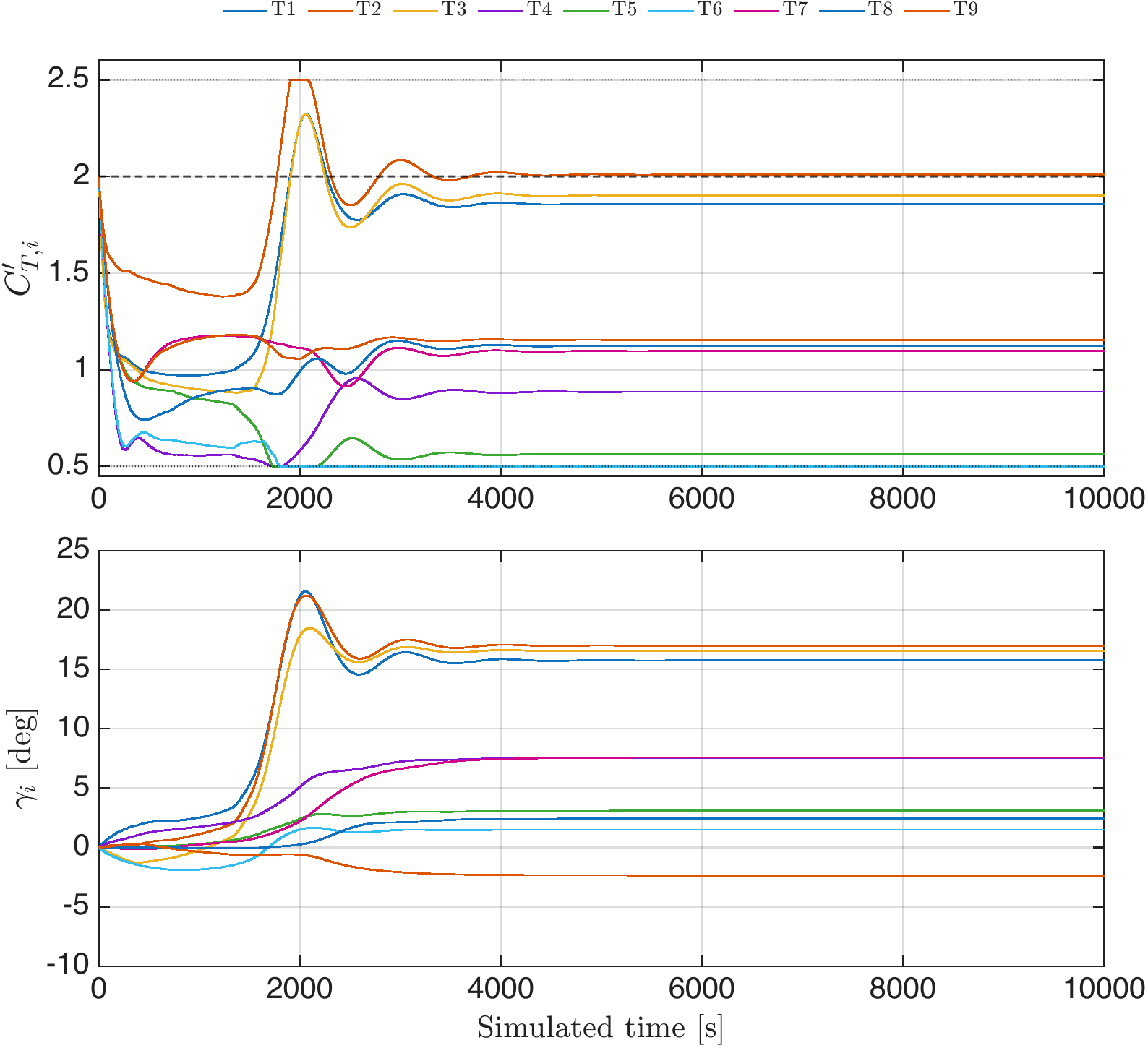}    
\caption{\textcolor{black}{SFO control trajectories for all nine turbines.}} 
\label{fig:sfo_control}
\end{center}
\end{figure}

\textcolor{black}{Fig. \ref{fig:sfo_control} reports the SFO control trajectories for all nine turbines. The disk-based thrust coefficients remain within the imposed constraints $C'_T\in[0.5,2.5]$. Although one trajectory briefly reaches the upper bound during the transient, the inputs subsequently move back to the interior of the admissible set. The yaw angles remain within the imposed $\pm30^\circ$ bound, with a maximum magnitude of approximately $21.6^\circ$. These trajectories indicate that the SFO gain is not obtained by sustained thrust saturation or by pushing the yaw angles to their imposed limits.}

\subsection{Computational Efficiency}
Computational tractability is a key challenge for full-farm closed-loop optimization, because accurate wake models are high-dimensional and coordination across many turbines leads to large optimization problems \citep{meyers2022wind}. Therefore, we assess the real-time feasibility of the three methods in this section. The presented results are obtained on an Apple M2 Pro and 16GB unified memory, using one core and MATLAB R2023b.

In terms of computational cost, ESC has the lowest online cost, with an average computation time of $0.007\ \mathrm{s}$ per control update. AMPC (RTI) reaches the reported power level in fewer simulation steps, but has the highest per-update cost, $20.76$ s, because each update requires forward prediction and adjoint-gradient computation. SFO requires $0.28$ s per update, which is much lower than AMPC (RTI), while its cumulative computation time to settle, $1927.0$ s, is comparable to that of AMPC (RTI), $2054.1$ s. For larger wind farms, the main additional cost of SFO is expected to come from the sensitivity calculation, which requires solving sparse linear systems of the form $(E(X_k)-\mathcal{A}_k)Z=B_k$. Since the WFSim state dimension grows with the number of grid cells, this cost is mainly affected by the larger sparse linear solves rather than by the turbine number alone.

\begin{figure}
\begin{center}
\includegraphics[width=8.8cm]{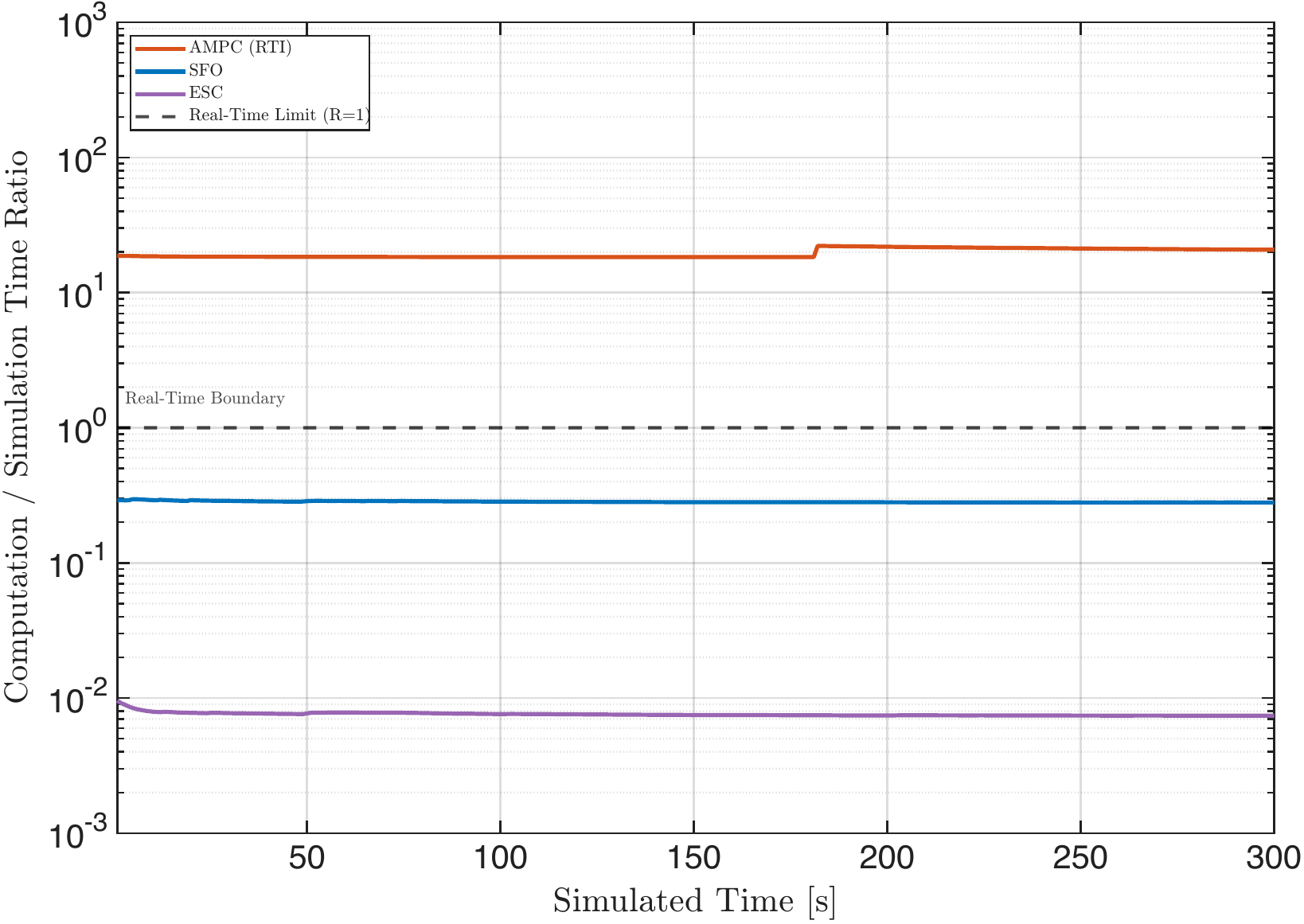}    
\caption{Ratio of cumulative computation time to simulated physical time (real-time feasibility metric).} 
\label{fig:real_time_feasibility}
\end{center}
\end{figure}

To further evaluate real-time feasibility, Fig. \ref{fig:real_time_feasibility} reports the ratio between the cumulative computation time and the simulated physical time $R_k:= \left(\sum_{j=1}^k t_j^{\mathrm{comp}}\right)/ t_k$, where $t_j^{\mathrm{comp}}$ is the measured computation time spent by the controller to produce the control applied at step $j$. Values below $R_k = 1$ indicate that control updates are computed faster than the plant evolves in simulation time. The results show that both SFO and ESC maintain this ratio below unity throughout the simulation, while AMPC (RTI) exceeds the real-time boundary, indicating its limitations for fast online implementation.

These comparisons show that SFO provides a balanced trade-off between power performance and computational efficiency. It achieves the highest steady-state power while remaining suitable for real-time control.   


\section{Conclusion}\label{sec:conclusion}

We compared three wind farm control strategies using the same WFSim setup with identical constraints and initialization. The results show that no single controller dominates in all aspects. Performance depends on the trade-off between steady-state power production, transient behavior, and available computational resources. In practice, the choice could be guided by which of these matters most for the application at hand. In our case study, SFO provides a reasonable balance under these conditions.

\textcolor{black}{The present benchmark is limited to a single steady-inflow scenario. A natural next step is to study time-varying wind speed and direction. In this case, the steady-state power-maximization problem becomes a time-varying optimization problem, where SFO should track a moving optimum rather than converge to a fixed operating point. Ideas from online feedback optimization for time-varying operating conditions \citep{colombino2019online} could be used to assess the tracking capability of the controllers. Further evaluation on larger wind farms and the inclusion of practical constraints such as structural loads and actuator activity are also of interest.}

\section*{DECLARATION OF GENERATIVE AI AND AI-ASSISTED TECHNOLOGIES IN THE WRITING PROCESS}
During the preparation of this work the authors used ChatGPT in order to improve the language and readability of the manuscript. After using this tool, the authors reviewed and edited the content as needed and take full responsibility for the content of the publication.

\bibliography{ifacconf}             
                                                   







\end{document}